# A PERSPECTIVE ON K-12 AI EDUCATION

Nathan Wang[1], Paul Tonko[2], Nikil Ragav[3], Michael Chungyoun[4], and Jonathan Plucker[5]

1. Department of Biomedical Engineering, Johns Hopkins University, Baltimore, MD, USA

2. United States House of Representatives, Washington, DC, USA

3. InventXYZ, Kansas City, MO, USA

4. Department of Chemical and Biomolecular Engineering, Johns Hopkins University, Baltimore, MD, USA

5. School of Education, Johns Hopkins University, Baltimore, MD, USA

Artificial intelligence (AI), which enables machines to learn to perform a task by training on diverse datasets, is one of the most revolutionary developments in scientific history. Although AI, and especially deep learning, is relatively new, it has already had a transformative impact on medicine, biology, transportation, entertainment, and beyond. As AI changes our daily lives at an increasingly fast pace, we are challenged with preparing our society for an AI-driven future. To this end, a critical step is to ensure an AI-ready workforce through education. Advocates of beginning instruction of AI basics at the K-12 level typically note benefits to the workforce, economy, and national security. In this complementary perspective, we discuss why learning AI is beneficial for motivating students and promoting creative thinking and how to develop a module-based approach that optimizes learning outcomes. We hope to excite and engage more members of the education community to join the effort to advance K-12 AI education in the United States and worldwide.

## INTRODUCTION

The modern timeline for artificial intelligence (AI) begins with the 1956 Dartmouth Summer Research Workshop, where eleven visionaries outlined both the significant challenges and promising directions in the field of AI. In the following decades, major trends synergized to elevate AI from theory to application, establishing data-driven approaches as a new paradigm for innovation.

One key trend is Moore's law, which projects a doubling in computing power every 24 months and has held empirically true since the late 20th century. By the early 2000s, the sustained exponential growth of computing power pushed the limits of computational complexity to a critical point, resulting in several performance breakthroughs in AI areas, such as object detection and natural language processing (1). In parallel, another major trend has been the rapid digitization of everyday goods and services into an Internet of Things, increasing the diversity and quantity of data available for analysis. In unison, these trends have developed into the data-driven paradigm that encompasses AI, machine learning (ML), and deep learning (DL). In general terms, AI refers to making computers do processes that humans normally do. Similarly, ML is a major subset of AI that involves computers learning from data and experience to perform a task. Furthermore, DL is a subset of ML that uses artificial neural networks, which are currently the mainstream of AI research (1,2). In this paper, we loosely refer to AI, ML, and DL as AI although it is important to know their nuances.

In contrast to traditional algorithms that produce an output based on a closed formula or an explicit computer program, data-driven algorithms learn to generate an optimal program from task-relevant data. This data is commonly organized as paired examples of input and expected output although semi-supervised and self-learning modes are becoming increasingly popular. The ability to learn and generalize makes AI suitable for complex tasks that are infeasible to address with conventional mathematical modeling and rules-based programming. Thus, AI has led to many impressive advances, including surgical robotics, self-driving vehicles, personalized entertainment, and automated diagnostic systems. The increasing momentum of AI can be clearly observed in Figure 1, which plots the rapid growth of AI research publications over time.

AI has established a firm foothold in science and industry and is now receiving serious attention from legislators. In January 2021, the landmark National Artificial Intelligence Initiative Act passed as a core section of the 2021 National Defense Authorization Act (3). Notably, this includes a provision for H.R. 8390, the AI Education Act, which was co-sponsored by the second author. The AI Education Act directs the National Science Foundation to establish competitive grants for the development of AI curricula at the K-12 level and for teacher training in science, technology, engineering, and mathematics (STEM) (4). As we deploy AI systems to accelerate innovation, improve our quality of life, and protect us from threats, we must take the critical step of making K-12 AI education a reality.

The political, economic, and scientific motivations for AI education are well-established (5). While these arguments are convincing, it is important that AI education does not feel like a burdensome obligation to educators – many of whom are already under workload pressures. Rather, this paper intends to introduce a new perspective: Learning and applying AI is a uniquely powerful way to motivate students and foster creativity in the classroom. These aims may be more relevant to K-12 teachers compared to the arguments about long-term societal benefits, which are somewhat removed from the classroom context.

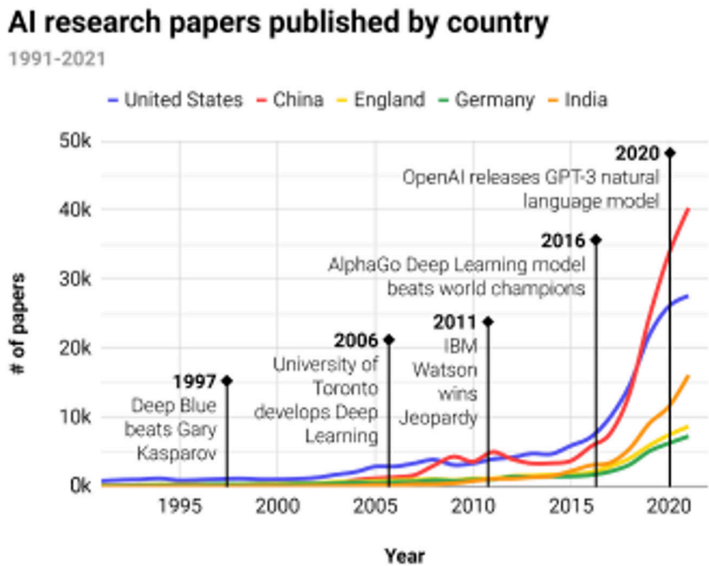

**Figure 1.** The number of AI publications over time. The data was obtained from Scopus using key words "artificial intelligence," "machine learning," or "deep learning."

In the following sections, we first explain how AI could succeed where the status quo fails in showing students the relevance and applicability of what they learn to the "real world." Then, we identify how working with AI provides students with an ideal environment for developing original, useful, and thus creative solutions. Finally, we provide examples of high school level AI project modules that teachers can effectively use and refine in their classes.

## LEARNING AI MOTIVATES LEARNING STEM

Although pursuing math and science for the sake of understanding can be an end in itself, many young students study STEM in order to apply what they are learning to the "real world." This expectation is reflected in one of the most common questions heard in the classroom: "When am I going to use this in life?" Fundamentally, they are really asking, "Why does this matter?" Providing a satisfying answer, given the time constraint of the class and knowledge gap of the student, is one of the inherent challenges for an educator. In the first author's experience as a recent high school graduate, this question is met with a combination of three answers:

Q: "When will I use this/why does it matter?"

A1: "Being proficient at math and science will lead to a successful career."

A2: "Here are some examples we looked at in class …"

A3: "You will understand how this is useful if you keep learning more."

The first response is generally true. Demand for jobs in STEM fields, which have some of the highest average salaries, is increasing. However, it is also true that 78% of the U.S. workforce exercises only basic addition, subtraction, multiplication, division, and fractions on a day-to-day basis (6). Thus, students who aspire to be lawyers, musicians, journalists, or other important non-STEM occupations understandably find it hard to see, for instance, how calculating a Taylor expansion relates to their fields of interest. In fact, even for students who are interested in a STEM career, the promise that what they are learning will be useful one day in the future is not likely to be what motivates them today. Thus, the challenge is to simultaneously provide students a clear idea of how they can use what they learn while they are learning it.

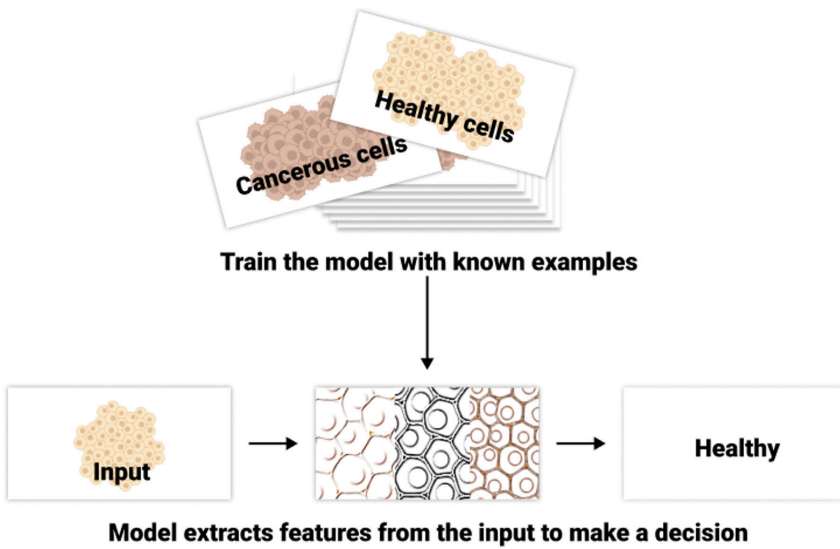

**Figure 2.** Training a supervised AI algorithm is analogous to a student studying for an exam with flashcards.

The second response attempts to cover the gaps in the first response with examples from classwork. For example, students in an advanced placement calculus class will be exposed to problems like calculating the volume of a revolved shape or maximizing the volume of a box given a constrained amount of cardboard (7). These applications are certainly interesting and valuable, but to the student, they may feel contrived and disappointingly unrealistic. The connection between solving these problems and inventing life-saving cures or transformative technologies is not made clear.

The third response continues from the second by explaining that the student needs to learn more in order to advance from solving classroom examples to solving complex real-world problems. Though we do not disregard that science research is built on years of cumulative knowledge, the concern is that if each time a student's curiosity is met with unsatisfying answers, he or she may become frustrated and lose motivation. As a result, the beauty, relevance, and underlying reason of math and science is lost in favor of rote memorization to pass exams. Ultimately, math and science lose their appeal as powerful tools and, instead, become an obligation and a chore.

The authors believe that learning AI equips students with a powerful and flexible set of tools and ways of thinking that would allow them to meaningfully explore realistic and significant real-world applications of what they study in class. Thus, we also suggest that working with AI has unique potential to motivate students in STEM subjects. To appreciate this point, the following four properties of AI and data-driven algorithms should be emphasized.

> **AI is powerful:** AI outperforms non-data-driven methods in numerous cases, especially on tasks that require high-level abstraction and that are too complicated to be explicitly modeled. Thus, AI is suitable for scenarios with a huge number of underlying variables that are best captured and represented in large datasets, such as a car navigating to a destination, detecting cancer in a medical scan, or predicting the 3D structure of proteins (8).
>
> **AI can be intuitive**: In a high-level sense, AI is inspired by human learning and behavior. The process of training a deep neural network, one of the most powerful AI algorithms today, can be compared to a student studying for a test with flashcards, as illustrated in Figure 2. For instance, a neural network designed to distinguish between cancerous and healthy tissues must be provided with many images of the two. The algorithm repeatedly runs through this stack of examples, tracking its own accuracy, adjusting its internal weighting, and optimizing its decision making.

At its core, this internal weighting is no more complex than integer multiplication and addition followed by a non-linear activation. A student with a background in algebra is well equipped to understand neural networks and AI at a reasonable level. Furthermore, the power of this straightforward flashcard analogy is that the functionality of the algorithm is not hidden behind opaque equations. Instead, each mathematical component of the neural network can be associated with a clear and intuitive purpose.

**AI is versatile:** AI can work with many types of inputs (image, text, audio, etc.) to produce many types of outputs (image, text, audio, etc.). As a consequence, AI algorithms can be readily adapted to solve highly diverse problems. For example, AI can analyze a medical image and output a diagnostic report, receive a noisy audio sample and extract individual voices, or accept a genetic sequence as text and identify potential genes. These three examples are shown in Figure 3. Understanding a few AI algorithms, such as a set of closely related neural networks, allows students to readily explore these applications and many more.

**AI can be accessible:** Teaching AI would not necessarily require expensive investment in computational infrastructure beyond what is already standard in a school district. A low-end laptop with internet access can take advantage of free cloud-based computing platforms, such as Google Colab, with access to numerous publicly available datasets (9). For high school students with no programming experience, visual AI development software, such as AI Studio, can reduce the barrier of entry. In AI Studio, the user can drag and drop code blocks and piece together their AI model as a flowchart, skipping confusing syntax and debugging for the time being (10,11). As the student becomes more comfortable, the built-in layers of abstraction can be peeled away until they are working directly with code.

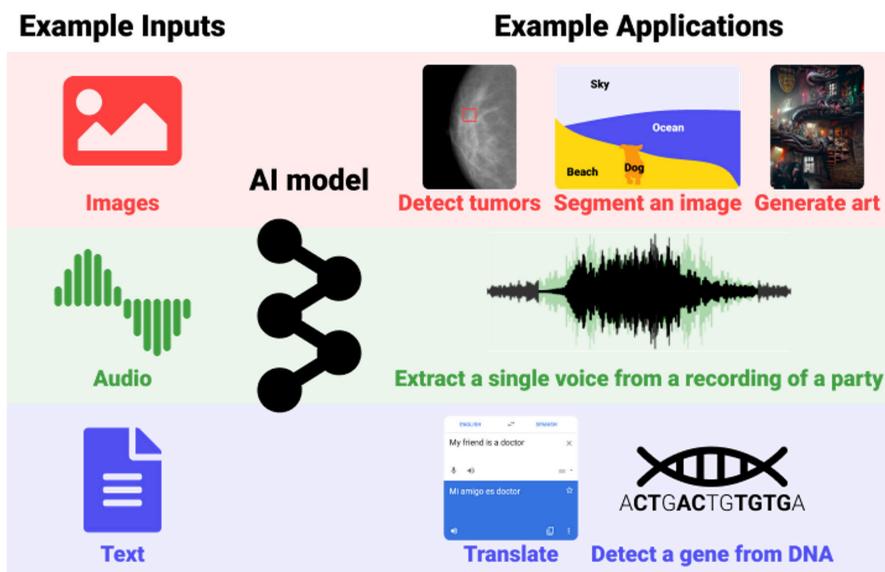

**Figure 3.** AI can process various types of inputs and produce good results in various applications, promising broad and lasting impacts on almost all scientific domains.

Putting these characteristics together, we see that AI is within reach for high school students, who can use it to build powerful solutions for a rich variety of realistic and interesting problems. These experiences could serve as a strong motivator for students in STEM classes by allowing learners to explore the relevance and usefulness of the topics they are learning.

## LEARNING AI FACILITATES CREATIVITY

In parallel to the challenge of keeping students motivated is the challenge of facilitating creativity in the classroom. A widely accepted definition of creativity is as follows: "Creativity is the interaction among aptitude, process, and environment by which an individual or group produces a perceptible product that is both novel and useful as defined within a social context"(12). This definition can be expressed as the equation C = [O x TA]context, where "C" stands for creativity, "O" stands for originality, "TA" stands for task appropriateness, and "context" represents the criteria for defining what is considered original and task appropriate (13). For example, a baking soda volcano may be creative in the context of an elementary school science fair but is not going to win accolades at a research conference. On the other hand, writing a poem about the weather in response to a question on an algebra test may be original but is not very task appropriate. Hence, we likely cannot say that it is creative.

For this discussion, we focus on how the activities and assignments in a class shape the context for creativity. For example, a math worksheet is not conducive to creativity since there is not much leeway to be original (you solve the problem, show your work, and write down the answer). Of course, this does not mean that worksheets have no value in schools at all, but having only worksheets for the entire year would be boring and would not maximize learning. In contrast, lessons that allow students to learn and apply AI form a context that is highly conducive to creativity. To demonstrate this point, we must be able to confidently say "yes" to the following two questions:

"Does AI make it easier for students to be original?" and "Is teaching AI task appropriate?"

Regarding the first question, there are many degrees of freedom inherent to designing an AI model. First, the student must engage in feature selection and data pre-processing, determining which sources of data are most relevant to their problem, how to format the data, and how to augment that data. Next, students can select from a variety of interrelated AI algorithms, mixing and matching the strengths of different models to design an optimal model such as through ensemble learning. Finally, the students can expand beyond the AI itself and look towards integrating their models into a broader solution. Take a student who has developed an algorithm to detect the sound of chainsaws in a protected forest. What are the next steps? How and where would the AI be deployed? Who gets alerted? How can this solution synergize with existing systems? These are all open-ended questions that have many valid approaches. In general, the capacity for originality is proportional to the multiplicity of potential design decisions. AI projects present students many unique avenues for solving a problem and thus make it easier to do something original.

Regarding the second question, recall that versatility was established as one of the key characteristics of AI. Hence, one can readily find relevant and task-appropriate AI applications in math, biology, chemistry, and physics. Thus, making time for AI projects in a STEM course is not an inappropriate excursion from class time but an effective enrichment activity that reinforces class topics with real-world applications. In addition, as mentioned in the preceding section, software tailored to K-12 students can drastically lower the learning curve for building AI models by eliminating the need to spend weeks on teaching programming syntax up front.

It is also possible for AI projects to be relevant in English and social studies classes. Students can "plug and play" with pre-built AI models, analyzing the outputs to explore the philosophical and ethical aspects of AI. For example, students in English can simulate the Turing test with the GPT-3 chatbot, analyzing the tone and sentence structures of responses to distinguish between human and machine. The point is that there are many possible examples of task-appropriate AI projects in the core subject areas, which give students greater room for creativity compared to writing another term paper, making another PowerPoint presentation, or filling out another worksheet.

## MODULE-BASED AI EDUCATION

Current forms of K-12 AI education in the United States tend toward one of two extremes (14). On one end, we have informal approaches, such as massive open online courses (MOOCs), extracurricular programs, and summer camps. On the other end, we see formal approaches, such as official AI elective courses, which are relatively less common. Many of the informal approaches, run by nonprofits, universities, and/or corporations, are of high quality and can rapidly adapt to new ideas. However, these programs are typically short-term, highly selective, and sometimes expensive. The exceptions are the free MOOCs available on websites like Coursera and edX (15), but these are generally designed for university students and above. MOOCs are appropriate for the self-motivated learner but may be too intimidating and difficult for many others. Similarly, AI electives are certified and fit well into a school district's STEM agenda but are rare and usually reserved for advanced learners attending a handful of elite high schools. Thus, we need to identify an intermediate approach to AI education that combines the advantages of informal and formal strategies.

In the previous sections, we examined key benefits of AI education on student motivation and creativity, noting that these benefits are best realized when AI concepts and projects are directly integrated into a student's regular coursework. Thus, we propose an intermediate approach to AI education that integrates hands-on AI project modules directly into the core subjects (math, science, English, and history). This would combine the adaptability of informal approaches with the necessary structure of a formal in-class setting while reaching all students in a given school. Alliances between schools and local universities, non-profits, and corporations can be forged to provide long-term support for teaching these modules and designing new ones, likely with federal grant support.

An example of such an alliance is represented by the collaboration between the first, fourth, and fifth authors, alongside other students at Johns Hopkins University, with the third author and his industry partners. The goal of this collaboration is to design AI modules for high schools in Baltimore and Kansas City. With the teachers and students in mind, the modules are split into four parts:

- Part 0: A review of course background material in the context of solving the problem presented

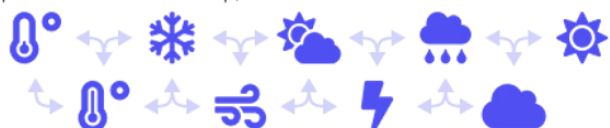
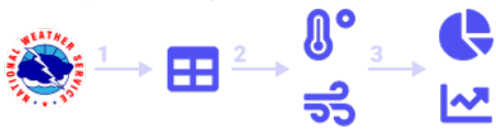
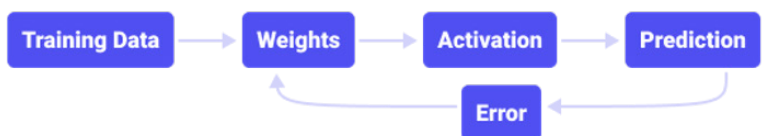

**Figure 4.** Weather forecast AI module.

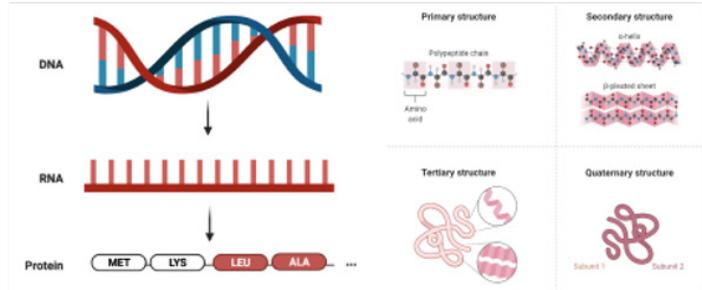
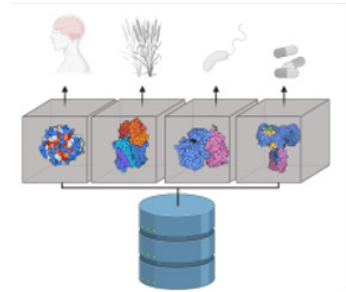
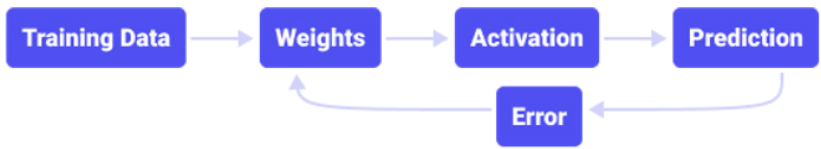
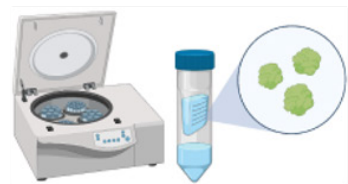

**Figure 5.** Molecular biology AI module.

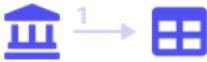
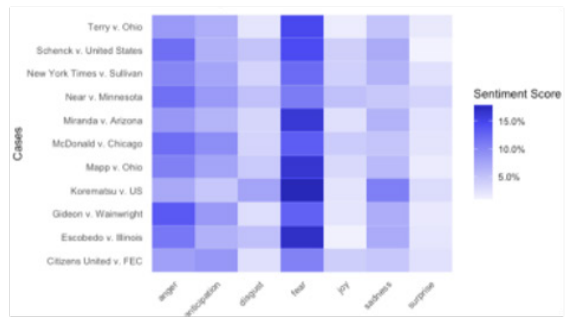
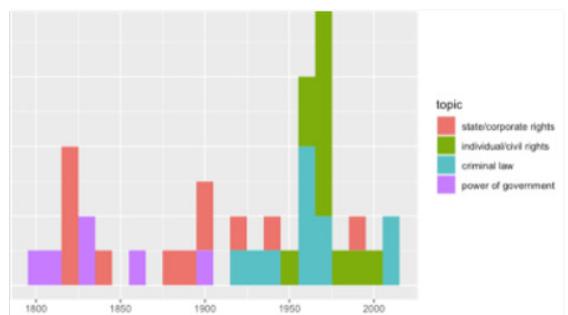

**Figure 6.** Analyzing Supreme Court cases using AI module.

in the module
- Part 1: Collection and exploration of relevant datasets through data visualization and feature engineering
- Part 2: Design and training of an AI model or experimenting with an existing model based on the data in Part 1
- Part 3: Application of the AI model to real-world situations to obtain tangible results

Under this scheme, a teacher has control over the amount of time allocated to each part. For instance, Part 2 can be condensed by providing more pre-written code to the student. Each part can also be completed independently, or the module can be divided between different classes.

One of the modules developed for a math or environmental science class is shown in Figure 4 and allows students to make their own weather forecast model. First, students retrieve data for their area from the National Weather Service, which includes precipitation, wind speed, humidity, UV index, daily weather conditions, and more. Next, students can pick and choose different variables and AI algorithms depending on the target they want to predict (e.g., a qualitative weather forecast vs. predicting the exact amount of snow). Finally, students have the option of writing a webapp in HTML and Python and hosting their model on the internet for anyone to use. Throughout this activity, students meaningfully apply important topics in weather, climate, and environment while simultaneously learning the fundamentals of AI.

Another one of our modules is for a biology class and allows students to train an AI on genomic data in order to predict the function of a gene from its base pair sequence as illustrated in Figure 5. During the course of the activity, students directly apply knowledge of the central dogma, amino acid interactions, protein structure, and more with neural networks. In the humanities, the modules shift focus from building AI algorithms to exploring the results of pre-built models. Figure 6 illustrates a module for students to spend time reading and analyzing Supreme Court cases. Students then observe the results of an AI algorithm clustering the cases by topic and identify how the trend of topics may correspond to different historical events and periods.

**CONCLUSION**

AI education presents a timely and exciting opportunity to motivate students and encourage them to think creatively. High school students, with the help of dedicated software tools, are capable of learning and applying AI. One promising approach to AI education is to integrate AI projects into the core subjects, giving students the freedom to develop original and powerful solutions while finding real-world applications of the topics they learn in their core classes. This approach centers on promoting alliances and partnerships among school districts, universities, nonprofits, and corporations.

Moving forward, there is a need for comprehensive and longitudinal data relevant to AI education to assess how the benefits, such as those described in this paper, can be best realized. This could be achieved through large-scale pilots across multiple school districts, which has not yet been done. We must also have important conversations about teacher training, professional development, and AI standards and certifications as well as about how AI and computer science agendas can complement each other (16).

There is a growing feeling of urgency and excitement regarding AI that first began with scientific and industry leaders and has now expanded to the national political stage. These sentiments provided the momentum, pushing the development of key AI technologies and policies. Now, we hope that this excitement will attract more colleagues in the education community, who are of critical importance in ensuring that the next generation of scientists, leaders, and citizens are prepared for the AI-driven future.

**ACKNOWLEDGMENTS**

The authors are grateful to Professor Rama Chellappa for providing his expertise in AI. The authors also thank the editors and anonymous reviewers for their detailed and helpful comments.